\documentclass[twocolumn,english,preprintnumbers,nofootinbib,showpacs,showkeys,prl]{revtex4}

\usepackage{amsmath,amsfonts,amssymb}
\usepackage{graphicx}
\usepackage[colorlinks=true, allcolors=blue]{hyperref}

\usepackage{bm}
\usepackage{mathrsfs}
\usepackage{feynmf}
\usepackage{verbatim}
\usepackage{tikz}
\usepackage{float}
\usepackage{subfigure}
\usepackage{xcolor}
\usepackage{float}

\begin{document} 


\title{Quantum computational representation of gauge field theory}

\author{Jeffrey Yepez}\email{yepez@hawaii.edu}
\date{October 17, 2016, revised December 18, 2017}
 \affiliation{
  Department\;of\;Physics\;and\;Astronomy,  University\;of\;Hawai`i\;at\;M\=anoa, Watanabe\;Hall,\;2505\;Correa\;Road, Honolulu,\;Hawai`i\;96822
}

\begin{abstract}
Presented is a  quantum computing model of a quantum field  theory for a system of fermions interacting via a massive gauge field.  The
model describes a relativistic superconducting fluid and uses a  metric tensor field to both encode the fermion's intrinsic spin in the torsion of curved space and encode the coupling of  fermions via a massive 4-potential field. The quantum computing model is a lattice model whose cell size is a deformation parameter: the equivalent lattice  and  curved-space gauge field theory models both reduce to quantum field theory in flat Minkowski space at zero cell size.  The low-energy expansions of the lattice model and Euler-Lagrange equations of the curved-space gauge field theory are the same equations of motion. The fermion and gauge fields obey the Dirac and Proca equations, and the gauge field strength is determined by the fermion field.  
\end{abstract}


\maketitle


{\it Introduction.}---Feynman started the field of quantum computation \cite{feynman-ces60,feynman-82}
 by suggesting that it takes one quantum system to efficiently simulate another---classical computers are less powerful.  He proposed a universal quantum simulator   \cite{feynman-85,divincenzo-pra95,divincenzo-pra95a,barenco-prsl95}:  a quantum system that can accurately simulate any other quantum system with resources that scale only linearly with the volume of  space  \cite{yepez_96_tech_report,bib:BB09,bib:BB07,bib:BB08,bib:BB06,RevModPhys.86.153}. Feynman also conjectured, separately, that a simulation with finite computational resources might be exact \cite{Feynman.1965}.  This letter confirms this foundational conjecture for a nontrivial model: a gauge field theory for a relativistic superconducting fluid with spin-1/2 fermions coupled via a massive spin-1 gauge field.

 Recently, the flood gates have opened in the search for lattice models of gauge field theories. 
Early approaches for fermions on a lattice---including loop algorithms 
 \cite{PhysRevB.50.136}, 
Kogut-Susskind (staggered) fermions
\cite{PhysRevD.91.054506} 
and 
Wilson fermions
\cite{PhysRevD.87.074505}---have led to new approaches including 
stream-collide approaches based on quantum walks
%
\cite{PhysRevA.94.032328,PhysRevA.93.052301,PhysRevA.94.012335,PhysRevLett.111.160602,PhysRevA.90.062106,Bisio2017}, digital lattice gauge theories with dynamical fermions
\cite{PhysRevA.95.023604,PhysRevA.95.042112,Jordan_2017} and tensor networks \cite{PhysRevX.4.041024}. 
Nambu-Jona-Lasinio models on a lattice are available
\cite{PhysRevD.86.034502}. 
Spinor quantum gravity
\cite{PhysRevD.85.104017,PhysRevD.86.104019}, 
spin foam
\cite{PhysRevD.80.064028,PhysRevD.83.025018}
and  liquids 
\cite{PhysRevB.89.235102,PhysRevB.94.224401} are undergoing active research.  
Supersymmetric quantum mechanics
\cite{PhysRevD.89.014511}, 
Wilson-loop discrete quantum gravity 
\cite{PhysRevD.81.084048}, 
and Yang-Mills theories 
\cite{PhysRevD.76.094503} are being actively investigated too. 
Finally, in the direction of quantum simulation, emulation using ultracold atoms (non-Abelian gauge field theory) is  undergoing rapid progress
\cite{PhysRevLett.110.055302,PhysRevA.88.023617,PhysRevA.94.063641,PhysRevLett.111.115303,PhysRevLett.110.125303,PhysRevLett.109.175302,PhysRevLett.109.145301,PhysRevLett.95.010404}. 
For Feynman-type quantum simulation, the new model presented here can be implemented as a unitary quantum algorithm on an array of qubits.

With the hope of revealing aspects of the yet-to-be-found theory of quantum gravity, one might also expect that the equivalent quantum computing model encode unitary particle and field dynamics using the metric tensor field of a curved-space manifold, along the lines of Mie's idea of an ``unavoidable connection" between gravitation and the existence of the fundamental particles \cite{Mie_1912a,Mie_1912b,Mie_1913}. The idea that space by itself can support all particle and field dynamics dates back to Kaluza \cite{Kaluza_1921}, Einstein \cite{einstein-1928b}  and Wheeler \cite{PhysRev.97.511,misner-73}, leading to string theory \cite{Green_Schwarz_Witten_1987_v1,Green_Schwarz_Witten_1987_v2} based only on a gravitational Lagrangian as a theory of everything.  Recently, Kempf has explored the idea that space can be simultaneously discrete and continuous \cite{PhysRevLett.103.231301,PhysRevLett.92.221301,PhysRevLett.85.2873,PhysRevD.71.023503,PhysRevD.71.023503,PhysRevD.69.124014,1367-2630-12-11-115001}.  Also, the idea that curved space is expressible in terms of spatial entanglement offers a quantum computational opportunity to emulate particle and field dynamics with a lattice model equivalent to a curved-space theory \cite{yepez:770202,Kempf2014,PhysRevLett.116.201101}.
%

 %
Applying Mie's approach to quantum information dynamics, the ansatz proposed here incorporates both gauge invariance and unitarity. 
This is realized by two propositions: (1) $g^{\mu\nu} = \eta^{\mu\nu} + \langle S^{\mu\nu} \rangle$ is the metric tensor field  and (2)   $\hbar \langle S^{\mu\nu} \rangle = -mc \ell \langle J^{\mu\nu} \rangle$ is the  quantization of torsion, where  $\eta^{\mu\nu}$ is the Minkowski metric,  $ \langle S^{\mu\nu} \rangle$ and  $\langle J^{\mu\nu} \rangle$ are expectation values of the angular momentum generators in the spin and position representations of the Lorentz group, $\hbar$ is the reduced Planck constant,  $m$ is the particle mass, $c$ is the speed of light,  $\ell$ is the cell size of the lattice.   From these propositions, gauge invariance and unitarity together are expressed as a forward finite-difference equation, giving  a lattice model  equivalent to a continuous curved space theory with the antisymmetric part of the metric tensor field $g^{\mu\nu}$ encoding torsion, a possibility 
  anticipated by Sciama  \cite{RevModPhys.36.463}.


{\it A curved-space gauge theory.}---A quantum computing model of  gauge  theory can be written as a generalization of flat-space quantum field theory where $g^{\mu\nu}$ 
models a unitary  $\psi$-$A^\mu$ nonlinear interaction
\begin{subequations}
\label{nonlinear_gauge_field_theory}
\begin{align}
{\cal L}
& =
 i \hbar c g^{\mu\nu}\overline{\psi}\gamma_\mu\left(\partial_{\nu}+\frac{ieA_\nu}{\hbar c}\right)\psi
-
 \frac{1}{4}F_{\mu\nu}F^{\mu\nu}
\label{nonlinear_Lagrangian_density}
,
\end{align}
where
$\psi
=
 (
   \psi_{\text{\tiny L} \uparrow}  ,
       \psi_{\text{\tiny L} \downarrow} ,
          \psi_{\text{\tiny R} \uparrow} ,
          \psi_{\text{\tiny R} \downarrow}
)^\text{\tiny T}
$
represents  fermions with mass $m$ and electric charge $e$.  The field operator's equal-time anticommutation relations are  $\{ \psi_s(\bm{x}), \psi_t(\bm{y})\}
 =0$, $\{ \psi_s^\dagger(\bm{x}), \psi_t^\dagger(\bm{y})\}=0$ and $
 \{ \psi_s(\bm{x}), \psi_t^\dagger(\bm{y})\}
 =\delta^{(3)}(\bm{x}-\bm{y})\delta_{st}$, 
   where $s$ and $t$ denote the spinor components of $\psi$. The fermions interact unitarily via a
 bosonic 4-potential field  $A^\mu = (A_0, \bm{A})$, where  $F^{\mu\nu} =\partial^\mu A^\nu - \partial^\nu A^\mu$ is the field tensor. The spacetime indices are $\mu,\nu = 0,1,2,3$.  
In the  $\ell = 0$ case the state of the fermion's intrinsic spin is encoded only in the 4-spinor field $\psi(x)$, while in the  $\ell \neq 0$ case its intrinsic spin is  self-consistently encoded in the asymmetric (torsional) part of $g^{\mu\nu}(x)$.
 The  quantization  condition 
 $\hbar \langle S^{\mu\nu} \rangle=-mc\ell\langle J^{\mu\nu} \rangle$ 
 equates
the expected value of the intrinsic spin (in units of $\hbar$) due to torsion
to
 the expectation value of the fermion's angular momentum (in units of  $mc\ell$)
  where  $S^{\mu\nu} = [\gamma^\mu, \gamma^\nu]/4$  and $J^{\mu\nu}= i\left( \ell \gamma^\mu  \partial ^\nu -\ell \gamma^\nu \partial ^\mu \right)$, and 
  where the expectation value of an operator $\hat O$ is defined as $\langle \hat O \rangle \equiv \overline{\psi}\hat O \psi/(\overline{\psi}\psi)$. 

Expanding the antisymmetric  $\psi$-dependent part of  $g^{\mu\nu}$  to first order in  $\epsilon = {  m_\circ c\ell }/\hbar$ gives
\begin{align}
\label{torsional_space_metric_tensor_field}
    g^{\mu\nu}(x)&=\eta^{\mu\nu}
    + i
    \frac{m_\circ c\ell}{\hbar}
    \frac{ \overline{\psi}(x) [\gamma^\mu, \gamma^\nu] \psi(x)}{\rho_\circ}
    +
    \cdots
    ,
\end{align}
\end{subequations}
 where  $m_\circ$ parametrizes the strength of the unitary interaction and $\rho_\circ$ is the   background  density 
 implicitly defined in the zeroth-order term in an $\epsilon$-expansion of $i\overline{\psi}\psi$
\begin{align}
\label{flux_expansion}
 \frac{4m_\circ c \ell}{\hbar}   i  \overline{\psi}(x)\psi(x) = \rho_\circ + \cdots 
    .
\end{align}

 The flat-space limit of (\ref{torsional_space_metric_tensor_field}) goes to  $g^{\mu\nu}(x)\xrightarrow{\ell \rightarrow 0} \eta^{\mu\nu}$,   so (\ref{nonlinear_Lagrangian_density})  reduces to the Lagrangian density for a quantum field theory with minimal coupling in continuous Minkowski space.  
For $\ell \neq 0$,  (\ref{nonlinear_gauge_field_theory}) 
is an analytical deformation of continuous quantum field theory that is congruent to a  quantum informational dynamics theory on a spacetime lattice with grid cell size $\ell$ and time step $\tau = \ell/c$.  $\ell\in[0,1]$
     is the deformation parameter of the theory.  The reason why  (\ref{nonlinear_gauge_field_theory})  is congruent to a discrete informational theory when  $\ell \neq 0$  is explained in this Letter; see (\ref{QID_equations_symmetrical_unitary_form}).

The    Euler-Lagrange equations obtained from (\ref{nonlinear_gauge_field_theory}) by varying $\psi$ and $A^\mu$ are
\begin{subequations}
\label{Euler_Lagrange_equations_nonlinear_GFT}
\begin{align}
\nonumber
i \hbar c g^{\mu\nu}  \gamma_\mu
\Big(
\partial_{\nu}&+\frac{ieA_\nu}{\hbar c}
\Big)
\psi
+
  i 2m_\circ c^2  S^{\mu\nu} \psi\,
\frac{  \overline{\psi}J_{\mu\nu}\psi }{\rho_\circ}
\\
\label{Euler_Lagrange_equations_nonlinear_GFT_a}
&=
 i \frac{e^2 \ell}{\hbar c}\left(-  \frac{m_\circ c^2}{e^2 \rho_\circ} e \overline{\psi}\gamma_\alpha \psi\right) A_\beta [\gamma^\alpha,\gamma^\beta]\psi
\\
\label{Euler_Lagrange_equations_nonlinear_GFT_b}
g^{\mu\nu} e \overline{\psi}\gamma_\mu \psi
&=
\partial_\mu F^{\mu\nu}
.
\end{align}
\end{subequations}
The Supplemental Material includes a derivation of (\ref{Euler_Lagrange_equations_nonlinear_GFT}).

{\it Relativistic superconductivity.}---A  relationship between the vector potential and charged current density was discovered by London, connecting Bose-Einstein condensates to nonrelativistic superfluidity and superconductivity \cite{PhysRev.54.947,PhysRev.74.562}.   A relativistic London relation  between the 4-potential  $A_\mu$ and    (unprimed)  4-current $J_\mu$ is
\begin{align}
\label{probability_current_solution_incoming}
 A_\mu(x) 
&= 
-  \frac{m_\circ c^2}{e^2 \rho_\circ} e \overline{\psi}(x)\gamma_\alpha \psi(x)
\equiv
-\lambda_\text{\tiny $L$}^2  e J_\mu(x)
,
\end{align}
where the squared London depth  is
$
\label{London_depth}
   \lambda_\text{\tiny L}^2\equiv {{m_\circ}c^2}/{(e^2\rho_\circ)}$, 
and minimizes action
 $\int d^4x\, {\cal L}$. The  righthand side of  (\ref{Euler_Lagrange_equations_nonlinear_GFT_a}) becomes $i
 ( {  e^2 \ell }/{(\hbar c) } )
A_\alpha  A_\beta [\gamma^\alpha, \gamma^\beta] \psi $ and vanishes because $[\gamma^\alpha, \gamma^\beta]$ is  antisymmetric  while  $A_\alpha  A_\beta$ is symmetric. 

Applying  quantization  $\hbar \langle S^{\mu\nu} \rangle = -mc \ell \langle J^{\mu\nu} \rangle$ of torsion and  the relativistic London relation (\ref{probability_current_solution_incoming}), as well as using the normalization condition  $\langle J^{\mu\nu} \rangle\langle J_{\mu\nu} \rangle =1$, 
 the equations of motion (\ref{Euler_Lagrange_equations_nonlinear_GFT_a})  and  (\ref{Euler_Lagrange_equations_nonlinear_GFT_b})  become  simpler
\begin{subequations}
\label{Yepez_Dirac_Maxwell_London_equations_of_motion}
\begin{align}
i \hbar c  \gamma^\nu
\Big(
\partial_{\nu}+\frac{ieA_\nu}{\hbar c}
\Big)
\psi
-
 m c^2 \psi
&=0
\label{Yepez_Dirac_Maxwell_London_equations_of_motion_a_lower_bound}
\\
\label{Yepez_Dirac_Maxwell_London_equations_of_motion_b_lower_bound}
g^{\mu\nu} e \overline{\psi}\gamma_\mu \psi
&=
\partial_\mu F^{\mu\nu}
\\
\label{Yepez_Dirac_Maxwell_London_equations_of_motion_c_lower_bound}
F^{\mu\nu} &=\partial^\mu A^\nu - \partial^\nu A^\mu
.
\end{align}
\end{subequations}
The Supplementary Material  has a derivation of (\ref{Yepez_Dirac_Maxwell_London_equations_of_motion}).
The definition of the field strength tensor  (\ref{Yepez_Dirac_Maxwell_London_equations_of_motion_c_lower_bound}) is added as a component equation. 
 These are relativistic equations of motion for a superconducting Fermi fluid, a generalization of the Dirac-Maxwell-London equations, that in the $\ell\rightarrow 0$ limit reduce to the 
Dirac-Maxwell equations  of quantum electrodynamics.

 {\it Forward and back reactions.}---The  4-current density on the lefthand side of (\ref{Euler_Lagrange_equations_nonlinear_GFT_b}) is $J'^\nu\equiv
g^{\mu\nu} \overline{\psi} \gamma_\mu \psi$, and this 4-current density is conserved
\begin{align}
\label{conserved_4_current}
\partial_\nu J'^\nu
\stackrel{(\ref{torsional_space_metric_tensor_field})}{=}
    \partial_\nu \left( \overline{\psi}\gamma^\nu \psi
  +
   i 
 \epsilon
 \frac{\overline{\psi}
[ \gamma^\mu ,
 \gamma^\nu
 ]
  \psi
}{\rho_\circ}\,
 \overline{\psi} \gamma_\mu \psi  
\right)=0 
.
\end{align}
This continuity equation follows from the  derivative of  (\ref{Euler_Lagrange_equations_nonlinear_GFT_b}), $\partial_\nu\partial_\mu F^{\mu\nu}=0$, because the derivatives are symmetric  under interchange of indices whereas the field tensor is antisymmetric.  The prime on  $J'^\nu$ denotes that the 4-current is an outgoing quantity 
\begin{subequations}
\label{outgoing_4_current}
\begin{align}
   J'^\mu 
&=g^{\mu\nu} e\overline{\psi} \gamma_\mu \psi
\\
\label{outgoing_4_current_b}
   &\stackrel{(\ref{torsional_space_metric_tensor_field})(\ref{probability_current_solution_incoming})}{=}
   e \overline{\psi}\gamma^\nu \psi
-i
 \frac{  e^2 \ell }{\hbar c } 
\overline{\psi} [\gamma^\mu, \gamma^\nu] \psi  
A_\mu
.
\end{align}
\end{subequations}
Applying the London relation (\ref{probability_current_solution_incoming}) again, the primed (outgoing) 4-potential is 
\begin{subequations}
\label{probability_current_solution_outgoing}
  \begin{align}
\label{probability_current_solution_outgoing_a}
-\frac{A'^\nu}{ \lambda_\text{\tiny L}^2}    
&\stackrel{(\ref{probability_current_solution_incoming})}{=}
J'^\nu
\\
\label{probability_current_solution_outgoing_b}
&\stackrel{(\ref{outgoing_4_current})}{=}
e \overline{\psi}\gamma^\nu \psi
-i
 \frac{  e^2 \ell }{\hbar c } 
\overline{\psi} [\gamma^\mu, \gamma^\nu] \psi  
A_\mu
.
\end{align}
\end{subequations}
 The primed and unprimed quantities are the incoming and outgoing states with respect to a unitary interaction.

\begin{figure}[!h!t!b!p]
\begin{center}
\subfigure[~QFT vertex]{\includegraphics[width=1.1in]{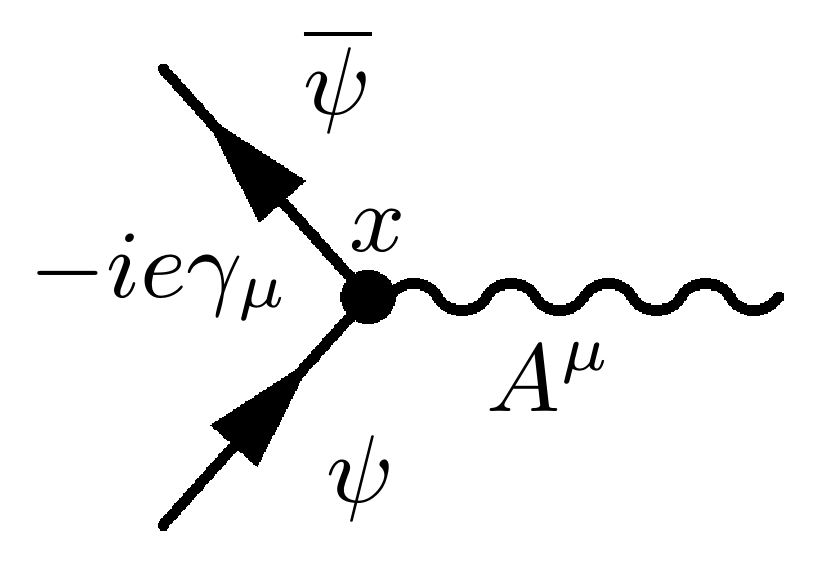}}
\subfigure[~QC unitary interaction]{\includegraphics[width=2.2in]{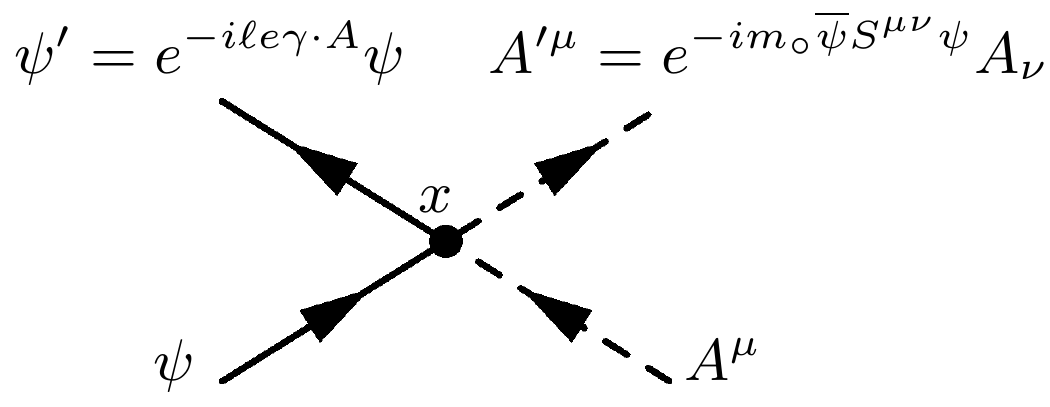}}
\caption{\label{Feynman_vertex_reaction} \footnotesize Comparison of $\psi$-$A^\mu$ interaction diagrams at a spacetime point $x$ in the (a) quantum field theory (QFT) and (b) quantum computing (QC) pictures.  The dashed lines in (b) depict the incoming fermion ${\cal A} = T \cdot A$ defined below in (\ref{4_spinor_A}) and outgoing fermion  ${\cal A}' = T \cdot A'$. The quantum computational unitary operators are written in natural units with $\hbar=1$ and $c=1$ and $S^{\mu\nu} = [\gamma^\mu, \gamma^\nu]/4$. The QFT vertex is useful for computing scattering amplitudes whereas the QC reaction $\psi(x) A^\mu(x)\rightarrow \psi'(x) A'^\mu(x)$ is useful for constructing an efficient quantum algorithm.}
\end{center}
\end{figure}

 It is helpful to rederive (\ref{probability_current_solution_outgoing}) in a different way to better comprehend the reason for the primed (outgoing) and unprimed (incoming) quantities.  ${\cal L}_\circ = i\hbar c \eta^{\mu\nu}\overline{\psi}\gamma_\mu\partial_{\nu}\psi$ represents the free particle motion and ${\cal L}'= -\overline{\psi}\gamma_\mu eA^\mu\psi$ represents the  minimal coupling $\psi$-$A^\mu$ interaction in the conventional flat-space gauge theory part of (\ref{nonlinear_Lagrangian_density}). The interaction in the flat-space gauge field theory is represented by a Feynman vertex at a point $x$,  depicted in Fig.~\ref{Feynman_vertex_reaction}a. 
Yet, in the equivalent quantum computing (QC) picture, the Feynman vertex diagram is just the low-energy representation of a unitary reaction depicted in Fig.~\ref{Feynman_vertex_reaction}b.  In the QC picture,   dynamics at the vertex (say driven by ${\cal L}_\circ + {\cal L}'\equiv\overline{\psi}L\psi$) is represented unitarily
\begin{equation}
\label{vextex_steady_state_condition}
\psi'(x)
=
e^{i  L \tau/\hbar} \psi(x) 
=
e^{- \ell  \gamma \cdot  
\left(\partial
 + i \frac{e A(x)}{\hbar c}
 \right) 
} 
\psi(x)
,
\end{equation}
where $\tau \equiv \ell/c$ is the update time.  The dynamics is conservative  when $\overline{\psi'}(x)\gamma^\mu\psi'(x)=\overline{\psi}(x)\gamma^\mu\psi(x)$ (Noether current), which allows one to define a condition for local equilibrium as
\begin{align}
\label{local_equilibrium}
A'^\mu(x) = A^\mu(x)
.
\end{align}
 The forward interaction part of  (\ref{vextex_steady_state_condition}),  where the incoming  $A^\mu(x)$ and $\psi(x)$ fields together produce the outgoing $\psi'(x)$ field, is therefore given by
\begin{equation}
\label{vertex_forward_reaction}
\psi'(x)
=e^{-i\ell  \gamma_\mu \frac{e A^\mu(x)}{\hbar c}}\psi(x) 
.
\end{equation}
Expanded to lowest order in $\ell$, (\ref{vertex_forward_reaction})  becomes
\begin{equation}
\label{psi_A_forward_reaction_expansion}
\psi'(x)
=
\psi(x) 
-
i e  \gamma_\mu \frac{\ell }{\hbar c} A^\mu(x) \psi(x)
+
\cdots.
\end{equation}
In this way, the second term on the righthand side of (\ref{psi_A_forward_reaction_expansion}) is the minimal-coupling interaction depicted  in Fig.~\ref{Feynman_vertex_reaction}a.  

There is also a back reaction (not explicitly encoded in (\ref{vextex_steady_state_condition})), whereby the incoming  $\psi(x)$ and $A^\mu(x)$ fields together produce the outgoing $A'^\mu(x)$ field. 
What is the equation for the back reaction? 
To answer this question, let us write  the outgoing 4-current on the righthand side of (\ref{probability_current_solution_outgoing_a}) in terms of the outgoing $\psi'$ field as 
\begin{align}
\label{incoming_charged_4_current}
   -\frac{A'^\nu(x)}{ \lambda_\text{\tiny L}^2} \stackrel{(\ref{probability_current_solution_incoming})}{=} e\overline{\psi'}(x)\gamma^\nu \psi'(x) 
   .
\end{align}
Inserting   (\ref{psi_A_forward_reaction_expansion}) into (\ref{incoming_charged_4_current}) gives $A'^\nu$ in terms of the incoming 4-potential $A^\nu$ and  $\psi$.  
  Making use of the adjoint gamma matrices $\gamma^{\mu\dagger}$ and  anticommutation relation $\{\gamma^{\mu\dagger}, \gamma^{\nu\dagger}\}=2\eta^{\mu\nu}$, the outgoing 4-current is
  \cite{yepez_arXiv1609.02225v2_quant_ph}
%
%
\begin{align}
\label{psi_A_back_reaction_expansion}
A'^\nu(x)
&\stackrel{(\ref{probability_current_solution_incoming})}{=}
A^\nu(x)
+
i 
  \frac{  {m_\circ}c\ell }{\hbar} 
\frac{\overline{\psi}(x)
[ \gamma^\mu ,
 \gamma^\nu
 ]
  \psi(x)
}{\rho_\circ}
A_\mu(x)
+
\cdots 
.
\end{align}
 The Supplemental Material  has a derivation of (\ref{psi_A_back_reaction_expansion}). 
This is just $A'^\nu(x)\stackrel{(\ref{torsional_space_metric_tensor_field})}{=}g^{\mu\nu}A_\mu(x)$,  so the outgoing field  is determined by  the  incoming  $\psi(x)$ field and the incoming $A^\nu(x)$ field.
The exact  flux-conserving collisional form of   (\ref{psi_A_back_reaction_expansion}) is a unitary update equation
\begin{equation}
\label{exact_back_reaction}
A'^\mu(x)
=
e^{
    i  
  \frac{  {m_\circ}c\ell }{\hbar} 
\frac{\overline{\psi}(x)
[ \gamma^\mu ,\gamma^\nu]
  \psi(x)}{\rho_\circ}
}
A_\nu(x)
.
\end{equation}
 The unitary interaction (\ref{exact_back_reaction}) preserves the norm $A'^\mu(x)A_\mu(x)=A^\mu(x)A_\mu(x)$ by driving $J^\nu=\eta^{\mu\nu}\overline{\psi} \gamma_\mu \psi$ to equal $J'^\nu=g^{\mu\nu}\overline{\psi} \gamma_\mu \psi$, which is consistent with (\ref{probability_current_solution_incoming}) and  (\ref{local_equilibrium}).  An advantage of the  unitary update (\ref{exact_back_reaction})  is that it provides a direct pathway to write $A^\mu$, and in turn the equations of motion (\ref{Yepez_Dirac_Maxwell_London_equations_of_motion}), in  4-spinor form \cite{yepez_arXiv1609.02225v2_quant_ph}.

{\it Gauge invariance.}---We require theory (\ref{nonlinear_Lagrangian_density}) be invariant under the gauge transformation
 \begin{align}
\label{gauge_transformation}
   \psi'
   &= 
   U \psi,
     \qquad
  A'^\nu
   = 
 U^\dagger A^\nu U
 -i \frac{\hbar c}{e} U^\dagger \partial^\nu U
 .
\end{align}
Yet, the alternative expressions (derived above) for $\psi'$ and $A'^\mu$ are  the unitary updates equations  (\ref{vertex_forward_reaction}) and (\ref{exact_back_reaction})
\begin{align}
\label{nonlinear_unitary_transformation}
    \psi'
=e^{-i\ell  \gamma_\mu \frac{e A^\mu}{\hbar c}}\psi
,
\qquad
    A'^\mu
=
e^{
    i  
  \frac{  {m_\circ}c\ell }{\hbar} 
\frac{\overline{\psi}
[ \gamma^\mu ,\gamma^\nu]
  \psi}{\rho_\circ}
}
A_\nu.
\end{align}
If $A^\mu$ is a pseudovector, then $\gamma_\mu A^\mu$ is an  hermitian matrix, and in turn the forward reaction equation (\ref{nonlinear_unitary_transformation}) is a unitary transformation.  
In the U(1) gauge theory case, there is freedom to invoke the gauge fixing condition $\chi =\ell  \gamma_\mu eA^\mu/(\hbar c)$, and inserting this into (\ref{gauge_transformation}) gives the simple update equations
 \begin{align}
\label{forward_difference_A}
\psi'=e^{-i \chi}\psi,
\quad
A'^\nu(x) 
=
A^\nu(x)
- \ell\gamma_\mu \partial^\nu A^\mu(x)
.
\end{align}
 Remarkably, the gauge transformation of $A^\mu$ becomes a forward finite-difference equation: continuous dynamics maps to discrete dynamics. The forward finite-difference (\ref{forward_difference_A}) offers a way to model (\ref{nonlinear_gauge_field_theory}) on a  lattice.  Inserting (\ref{nonlinear_unitary_transformation}) into (\ref{forward_difference_A}) gives a 4-vector Helmholtz equation
 \begin{align}
\label{wave_equation_4_vector}
    \partial^2 A^\mu   
      +
\frac{1}{ \lambda_\text{\tiny L}^2} g^{\mu\nu} A_\nu
=
0, 
\end{align}
 the Proca equation in equilibrium (\ref{local_equilibrium}) with $\partial_\mu A^\mu = 0$. 
  Inserting  (\ref{psi_A_back_reaction_expansion}) into (\ref{forward_difference_A}) leads to $F^{\mu\nu}=      
\partial^\nu A^\mu
-
      \partial^\mu A^\nu=
F^{\mu\nu}
=
\left.
 i \frac{\epsilon}{4 \ell\rho_\circ}
 \middle(
 \gamma^\mu
\overline{\psi}
[ \gamma^\lambda ,
 \gamma^\nu
 ]
  \psi
-
 \gamma^\nu
\overline{\psi}
[ \gamma^\lambda ,
 \gamma^\mu
 ]
  \psi
 \right)
A_\lambda
 +
 \cdots$. 

 {\it Quantum computational spinor form.}---
 As a warmup for rewriting the equations of motion (\ref{Yepez_Dirac_Maxwell_London_equations_of_motion_b_lower_bound}) and (\ref{Yepez_Dirac_Maxwell_London_equations_of_motion_c_lower_bound}) in spinor form, the Maxwell equations can be written in spinor form   \cite{yepez_arXiv1609.02225v2_quant_ph}, a generalization of the representation by Laporte and Uhlenbeck  \cite{PhysRev.37.1380}. Start by converting the contravariant 4-potential $A^\mu=(A_0, A_x, A_y, A_z)^\text{T}$
  into the 4-spinor field, say ${\cal A}$, by using a unitary matrix transformation, say $T$.  The unitary transformation is 
 \begin{align}
\label{4_spinor_A}
{\cal A}_a =  T_{a\mu} A^\mu
.
\end{align}
Defining $F^\mu\equiv(-\partial\cdot A, -\partial_0\bm{A}-\partial_0 A_0+ i \nabla\times \bm{A})$, 
a  4-spinor electromagnetic  field $\tilde {\cal F}$,   current density spinor  field ${\cal J}$,  and  dual 4-potential spinor field $\tilde {\cal A}$  respectively are 
\begin{equation}
\label{4_spinor_field_definititions}
\tilde {\cal F }
 =  
 T_{a\mu}F^\mu
 ,
\quad
{\cal J }
 = 
 T_{a\mu}J^\mu
,
\quad
\tilde {\cal A}
=
-i \lambda_\text{\tiny L}
 T_{a\mu}F^\mu
.
\end{equation}
The Maxwell equations 
  expressed in terms of 4-spinor fields (\ref{4_spinor_field_definititions}) and using tensor-product notation  are \cite{yepez_arXiv1609.02225v2_quant_ph}
\begin{align}
\label{Maxwell_equation_4spinor_rep_covariant_form}
e {\cal J} +  \bm{1}\otimes \sigma\cdot \partial \tilde {\cal F} &=
0,
\qquad
\tilde {\cal F} +  \bm{1}\otimes   \bar\sigma\cdot \partial {\cal A}=
0,
\end{align}
where
$\sigma^\mu = (1,\bm{\sigma})$, $\bar\sigma^\mu=(1,-\bm{\sigma})$.   
The Supplemental Material  contains a derivation of (\ref{Maxwell_equation_4spinor_rep_covariant_form}). 

Let us now write   (\ref{Yepez_Dirac_Maxwell_London_equations_of_motion}) in spinor form. 
Define a tensor field 
${M^\nu}_\mu(x) \equiv i m_\circ 
   {\overline{\psi}(x)
[\gamma^\nu,\gamma_\mu]
  \psi(x)}/{\rho_\circ}$. 
Then (\ref{outgoing_4_current_b}) is
\begin{subequations}
\begin{align}
eJ'^\nu(x)
=
-\frac{A^\nu(x)}{ \lambda_\text{\tiny L}^2}
+
 \frac{c\ell}{\hbar \lambda_\text{\tiny L}^2}
{M^\nu}_\mu(x)
A^\mu(x)
.
\end{align} 
\end{subequations}
This becomes
$
   e{\cal J}
\stackrel{(\ref{4_spinor_A})}{=}
-
\frac{1}{ \lambda_\text{\tiny L}^2}
\left(
1
-
 \frac{ T M T^\dagger c \ell}{\hbar}
\right)
{\cal A}
$, 
so the spinor representation of 
 (\ref{Yepez_Dirac_Maxwell_London_equations_of_motion}) 
  is
\begin{subequations}
\label{nonlinear_Maxwell_equation_4spinor_rep_covariant_form}
\begin{align}
i \hbar c  \gamma^\nu
\Big(
\partial_{\nu}+\frac{ieA_\nu}{\hbar c}
\Big)
\psi
-
 m c^2 \psi
&=0
\\
\label{nonlinear_Maxwell_equation_4spinor_rep_covariant_form_b}
-
\frac{1}{ \lambda_\text{\tiny L}^2}
\left(
1
-
 \frac{  T M T^\dagger c \ell}{\hbar}
\right)
{\cal A}
+ 
 \bm{1}\otimes \sigma\cdot \partial \tilde {\cal F} &=
0
\\
\label{nonlinear_Maxwell_equation_4spinor_rep_covariant_form_a}
\tilde {\cal F} +  \bm{1}\otimes   \bar\sigma\cdot \partial {\cal A}=
0
.
\end{align}
\end{subequations}
The equation pair (\ref{nonlinear_Maxwell_equation_4spinor_rep_covariant_form}) can be solved for ${\cal A}$, leading to the second-order  equation for a massive gauge field
   $-\partial_\mu \partial^\mu {\cal A} 
   -
\frac{1}{ \lambda_\text{\tiny L}^2}
\left(
1
-
 \frac{T M T^\dagger c \ell}{\hbar}
\right)
{\cal A}
=0$, 
 which in 4-vector notation is (\ref{wave_equation_4_vector}). 
Since $\tilde {\cal F} \stackrel{(\ref{4_spinor_field_definititions})}{=} i  \tilde {\cal A}/\lambda_\text{\tiny L}$, 
 (\ref{nonlinear_Maxwell_equation_4spinor_rep_covariant_form})  may be written in component form for single and doublet spinors
\begin{subequations}
\label{nonlinear_superconductivity_equation_spinor_form}
\begin{align}
\label{nonlinear_superconductivity_equation_spinor_form_a}
\begin{pmatrix}
  - \frac{mc}{\hbar}  &  i \sigma \cdot \left(\partial +i \frac{e A}{\hbar c}\right) \\
  i \bar{\sigma} \cdot \left(\partial+i\frac{eA}{\hbar c}\right)    & - \frac{mc}{\hbar}
\end{pmatrix}
\begin{pmatrix}
      \psi_\text{\tiny L}    \\
       \psi_\text{\tiny R}  
\end{pmatrix}
&\stackrel{(\ref{Yepez_Dirac_Maxwell_London_equations_of_motion_a_lower_bound})}{=}
 0
\\
\label{nonlinear_superconductivity_equation_spinor_form_b}
\begin{pmatrix}
  - \frac{1}{\lambda_\text{\tiny $L$}}
   \left(
1
-
 \frac{T M T^\dagger  c \ell}{\hbar}
\right)
   &  i \bm{1}\otimes\sigma \cdot  \partial    \\
  i  \bm{1}\otimes\bar{\sigma} \cdot \partial    &  - \frac{1}{\lambda_\text{\tiny $L$}}
\end{pmatrix}
\begin{pmatrix}
      {\cal A}    \\
     \tilde {\cal A}
\end{pmatrix}
&\stackrel{(\ref{nonlinear_Maxwell_equation_4spinor_rep_covariant_form})}{=}
 0
 .
\end{align}
\end{subequations}
These are a generalization of the Dirac-Maxwell-London equations derived in the Supplemental Material. 

Equation of motion  (\ref{nonlinear_superconductivity_equation_spinor_form_b})  are a Dirac equation  for a doublet field  (a pair of 4-spinor fields)  that has a block diagonal mass matrix
\begin{align}
\label{M_L_formula}
  M_\text{\tiny $L$} 
    &\equiv
        \frac{\hbar}{\lambda_\text{\tiny $L$} c} 
    \begin{pmatrix}
1
-
 \frac{T M T^\dagger c \ell}{\hbar}
 &
 0
      \\
0&    1 
\end{pmatrix}
=
        \frac{\hbar}{\lambda_\text{\tiny $L$} c} 
-
\frac{\ell}{\lambda_\text{\tiny $L$}}
    \begin{pmatrix}
T M T^\dagger
 &
 0
      \\
0&    0 
\end{pmatrix}
.
\end{align}
Rescaling the doublet field in (\ref{nonlinear_superconductivity_equation_spinor_form_b}) as 
\begin{equation}
\Phi = 
\begin{pmatrix}
      \Phi_\text{\tiny upper}    \\
       \Phi_\text{\tiny lower} 
\end{pmatrix}
=
 \frac{1}{\lambda_\text{\tiny $L$}} \sqrt{\frac{\tau}{\hbar}}
\begin{pmatrix}
      {\cal A}    \\
      \tilde {  \cal A}
\end{pmatrix} ,
\end{equation}
(\ref{nonlinear_superconductivity_equation_spinor_form}) may be compactly written as 
\begin{subequations}
\label{Dirac_Maxwell_London_equations_symmetrical_4_vector_form}
\begin{align}
\label{Dirac_Maxwell_London_equations_symmetrical_4_vector_form_a}
i\hbar c \,\gamma_\mu \left(\partial^\mu  + i \frac{e A^\mu}{\hbar c}\right) \psi
-
     m c^2\psi
 &=
 0
  \\
\label{Dirac_Maxwell_London_equations_symmetrical_4_vector_form_b}
i\hbar c \,{\cal G}_\mu
 \partial^\mu  
\Phi
-
      M_\text{\tiny $L$} c^2 \Phi
     &=
 0
,
\end{align}
\end{subequations}
where ${\cal G}^\mu=({\cal G}_0, \bm{{\cal G}})$ is a generalized Dirac 4-vector that  in a chiral $8\times 8$ matrix representation has  components $
{\cal G}_0 = \sigma_x\otimes \bm{1} \otimes\bm{1}
$
 and
 $
\bm{{\cal G}} = i\sigma_y \otimes\bm{1} \otimes\bm{\sigma}$.

A unitary representation of (\ref{Dirac_Maxwell_London_equations_symmetrical_4_vector_form})  on  
a spacetime lattice  has the  unitary forward and back 
reactions  (\ref{nonlinear_unitary_transformation}) contained within the particle and gauge field dynamics
\begin{subequations}
\label{QID_equations_symmetrical_unitary_form}
\begin{align}
\psi(x)
 &=
e^{- \ell  \gamma \cdot  
\left(\partial
 + i \frac{e  A(x) }{\hbar c}
 \right) 
- i  \frac{  m c \ell}{\hbar}  
} 
\psi(x)
\label{QID_equations_symmetrical_unitary_form_psi_evolution}
    \\
\Phi(x)
 &=
e^{-\ell  {\cal G}
 \cdot 
 \partial
- i  \frac{  M_\text{\tiny $L$}(x) c \ell}{\hbar}
}\Phi(x) .
\qquad 
\label{QID_equations_symmetrical_unitary_form_Phi_evolution}
\end{align}
\end{subequations}
  The small $\ell$ (low-energy) expansions of (\ref{QID_equations_symmetrical_unitary_form})
are  Euler-Lagrange equations  (\ref{Dirac_Maxwell_London_equations_symmetrical_4_vector_form}).\footnote{The curved-space gauge field theory (\ref{nonlinear_gauge_field_theory}) is approximated  
in the mean-field limit  by a flat-space gauge theory for    $\psi$ and 
    $\Phi$ 
\begin{equation*}
\label{fermionic_superconducting_Lagrangian_denstity_minimal_coupling}
\begin{split}
{\cal L}  
&
 =
 i \hbar c \overline{\psi}\gamma_\mu\left(\partial^{\mu}+\frac{i e}{\hbar c} \langle A^\mu\rangle \right)\psi
 -   m c^2 \overline{\psi}\psi
 \\
 &
  +
 i \hbar c  \overline{\Phi}{\cal G_\mu}
 \partial^{\mu}
 \Phi
 -  \langle M_\text{\tiny $L$}\rangle c^2 \overline{\Phi}\Phi
,
 \end{split}
\end{equation*}
 where $\langle A^\mu\rangle$ is a classical field and $\langle M_\text{\tiny $L$}\rangle$ is a mass matrix. 
}
 The Supplemental Material outlines a quantum  algorithm based on model (\ref{QID_equations_symmetrical_unitary_form}).

{\it Conclusion.}---The choice of using the curved-space quantum field theory (\ref{nonlinear_gauge_field_theory}) to formulate the quantum computing model (\ref{QID_equations_symmetrical_unitary_form}) was not  a choice made \`a priori. Instead, requiring that the  fermion-gauge field interactions be unitary transformations  and requiring that the Lagrangian density be gauge invariant forces this choice upon us.  
 So  quantum computation served as a pathway to discover a model quantum field theory  in curved space (where   the torsion of space encodes the particle's intrinsic spin) that is equivalent to a unitary lattice model.  Both models describe a gauge field theory of a superconducting Fermi fluid.

{\it Acknowledgement.}---I would like to thank Professor Xerxes Tata  for his helpful comments on this work and especially his advice to make sure the theory is gauge theory. I also thank Dr. Norman Margolus for helpful discussions about this new theory and presentation. This research was supported by the grant  ``Quantum Computational Mathematics for Efficient Computational Physics"  from the Air Force Office of Scientific Research.

\newpage
\section{.}
\vspace{10in}
\newpage

\section{Supplemental Material}



\subsection{Nonlinear gauge field theory}

\subsubsection{A curved-space gauge theory}

Consider the gauge-invariant Lagrangian density as a quantum information dynamics theory of superconductivity
\begin{align}
\label{SM_nonlinear_GFT_lagrangian_density}
{\cal L}
& =
 i \hbar c g^{\mu\nu}
 \overline{\psi}\gamma_\mu\left(\partial_{\nu}+\frac{ieA_\nu}{\hbar c}\right)\psi
-
 \frac{1}{4}F_{\mu\nu}F^{\mu\nu}
\end{align}
with matter field 
$\psi
=
 (
   \psi_{\text{\tiny L} \uparrow}  ,
       \psi_{\text{\tiny L} \downarrow} ,
          \psi_{\text{\tiny R} \uparrow} ,
          \psi_{\text{\tiny R} \downarrow}
)^\text{\tiny T}
$
for fermions with electric charge $e$ and mass $m$, 
  4-potential field  $A^\mu = (A_0, \bm{A})$ and field tensor $F^{\mu\nu} =\partial^\mu A^\nu - \partial^\nu A^\mu$,
   where $\mu,\nu = 0,1,2,3$.
  The metric tensor is
\begin{align}
\label{SM_torsional_space_metric_tensor_field_Smunu_form}
    g^{\mu\nu}(x)&=\eta^{\mu\nu}
    + 
    \frac{ \overline{\psi}(x)\frac{1}{4} [\gamma^\mu, \gamma^\nu] \psi(x)}{\overline{\psi}(x) \psi(x)}
    .
\end{align}
The zeroth-order term in an $\epsilon$-expansion of the flux $i\overline{\psi}\psi$ implicitly defines the   background number density $\rho_\circ$  
\begin{align}
\label{SM_flux_expansion}
 \frac{4m_\circ c \ell}{\hbar}   i  \overline{\psi}(x)\psi(x) = \rho_\circ + \cdots 
    ,
\end{align}
 so the metric tensor is
\begin{align}
\label{SM_torsional_space_metric_tensor_field_form_2}
    g^{\mu\nu}(x)&=\eta^{\mu\nu}
    + i
    \frac{m_\circ c\ell}{\hbar}
    \frac{ \overline{\psi}(x) [\gamma^\mu, \gamma^\nu] \psi(x)}{\rho_\circ}
    +
    \cdots
    ,
\end{align}
 where  $m_\circ$ parametrizes the strength of the nonlinear interaction. 
  The quantity $\ell$ denotes the smallest length scale.

  The Euler-Lagrange equations are obtained by minimizing the action $S=\int d^4x\,{\cal L}_\text{\tiny QID}$  with respect to  variations in $\psi$ and $A^\mu$ 
\begin{subequations}
\label{SM_Euler_Lagrange_equation}
\begin{align}
\partial_\mu\left(\frac{\partial {\cal L}[\psi,A]}{\partial(\partial_\mu \psi)}\right)
-
\frac{\partial{\cal L}[\psi,A]}{\partial \psi}
&=
0
\\
\partial_\mu
\left(
\frac{\partial{{\cal L}[\psi,A]}}{\partial(\partial_\mu A_\nu)} 
\right)
-
\frac{\partial{{\cal L}[\psi,A]}}{\partial A_\nu}
&=0
.
\end{align}
\end{subequations}
Inserting (\ref{SM_nonlinear_GFT_lagrangian_density}) into (\ref{SM_Euler_Lagrange_equation}) gives the  set of coupled  equations  
   \begin{widetext}
\begin{subequations}
\label{SM_Euler_Lagrange_equations_nonlinear_GFT}
\begin{align}
i\hbar c g^{\mu\nu}\gamma_\mu \left(\partial_\nu  - i \frac{eA_\nu}{\hbar c}\right) \psi
-
g^{\mu\nu} e \overline{\psi} 
\gamma_\mu\psi \frac{\partial A_\nu}{\partial  \overline{\psi} }
-
 \frac{1}{2}F_{\mu\nu}\frac{\partial F^{\mu\nu}}{\partial  \overline{\psi} }
+
i\hbar c \frac{\partial g^{\mu\nu}}{\partial  \overline{\psi} }
 \overline{\psi}\gamma_\mu \left(\partial_\nu  - i \frac{eA_\nu}{\hbar c}\right) \psi 
&= 0
\\
 g^{\mu\nu} e \overline{\psi}\gamma_\mu \psi
 +
i\hbar c \frac{\partial g^{\mu\sigma}}{\partial  A_\nu} \overline{\psi}\gamma_\mu\left(\partial_\sigma  - i \frac{eA_\sigma}{\hbar c}\right) \psi 
&=
\partial_\mu F^{\mu\nu}
.
\end{align}
\end{subequations}
In theory (\ref{SM_nonlinear_GFT_lagrangian_density}), $\psi$ and $A^\mu$ are considered to be independent fields, so the Euler-Lagrange equations reduce to
\begin{subequations}
\label{SM_Euler_Lagrange_equations_nonlinear_GFT_2}
\begin{align}
i\hbar c g^{\mu\nu}\gamma_\mu \left(\partial_\nu  - i \frac{eA_\nu}{\hbar c}\right) \psi
+
i\hbar c \frac{\partial g^{\mu\nu}}{\partial  \overline{\psi} }
 \overline{\psi}\gamma_\mu \left(\partial_\nu  - i \frac{eA_\nu}{\hbar c}\right)\psi 
&= 0
\\
 g^{\mu\nu} e \overline{\psi}\gamma_\mu \psi
&=
\partial_\mu F^{\mu\nu}
.
\end{align}
\end{subequations}
The variation of the metric tensor is
\begin{align}
\label{SM_torsional_space_metric_tensor_field_form_2}
 \frac{\partial g^{\mu\nu}}{\partial  \overline{\psi} }
 &=
   i \frac{m_\circ c\ell}{\hbar \rho_\circ}
[\gamma^\mu, \gamma^\nu] \psi
    +
    \cdots
    ,
\end{align}
 so the Euler-Lagrange equations further reduce to
\begin{subequations}
\label{SM_Euler_Lagrange_equations_nonlinear_GFT_3}
\begin{align}
i\hbar c g^{\mu\nu}\gamma_\mu \left(\partial_\nu  - i \frac{eA_\nu}{\hbar c}\right) \psi
-
\frac{m_\circ c^2\ell }{\rho_\circ}[\gamma^\mu, \gamma^\nu] \psi\,
 \overline{\psi}\gamma_\mu \left(\partial_\nu  - i \frac{eA_\nu}{\hbar c}\right)\psi 
&= 0
\\
 g^{\mu\nu} e \overline{\psi}\gamma_\mu \psi
&=
\partial_\mu F^{\mu\nu}
,
\end{align}
\end{subequations}
or
\begin{subequations}
\label{SM_Euler_Lagrange_equations_nonlinear_GFT_3}
\begin{align}
i\hbar c g^{\mu\nu}\gamma_\mu \left(\partial_\nu  - i \frac{eA_\nu}{\hbar c}\right) \psi
+
i \frac{m_\circ c^2}{\rho_\circ} [\gamma^\mu, \gamma^\nu] \psi\,
 \overline{\psi}(i  \ell\gamma_\mu \partial_\nu )\psi 
-
  i \frac{m_\circ c^2\ell e}{\hbar c\rho_\circ} [\gamma^\mu, \gamma^\nu] \psi\,
 \overline{\psi}\gamma_\mu  A_\nu\psi
&=
0
\\
 g^{\mu\nu} e \overline{\psi}\gamma_\mu \psi
&=
\partial_\mu F^{\mu\nu}
.
\end{align}
\end{subequations}
This can be rewritten as
\begin{subequations}
\label{SM_Euler_Lagrange_equations_nonlinear_GFT_4}
\begin{align}
i\hbar c g^{\mu\nu}\gamma_\mu \left(\partial_\nu  - i \frac{eA_\nu}{\hbar c}\right) \psi
+
i \frac{m_\circ c^2}{\rho_\circ} [\gamma^\mu, \gamma^\nu] \psi\,
 \overline{\psi}(i  \ell\gamma_\mu \partial_\nu )\psi 
+
  i  \frac{ \ell e^2}{\hbar c}[\gamma^\mu, \gamma^\nu]
  \psi \left(
 - \frac{m_\circ c^2 }{e^2\rho_\circ}  e \overline{\psi}
\gamma_\mu  \psi
\right)
A_\nu
&=
0
\\
 g^{\mu\nu} e \overline{\psi}\gamma_\mu \psi
&=
\partial_\mu F^{\mu\nu}
.
\end{align}
\end{subequations}
The  generator of angular momentum in the position-representation  of the Lorentz group is
\begin{align}
\label{SM_position_representation_angular_momentum_operator}
   J^{\mu\nu}
   &= i\left( \ell \gamma^\mu  \partial ^\nu -\ell \gamma^\nu \partial ^\mu \right) ,
\end{align}
so (\ref{SM_Euler_Lagrange_equations_nonlinear_GFT_4}) may be written as
\begin{subequations}
\label{SM_Euler_Lagrange_equations_nonlinear_GFT_5}
\begin{align}
i\hbar c g^{\mu\nu}\gamma_\mu \left(\partial_\nu  - i \frac{eA_\nu}{\hbar c}\right) \psi
+
i \frac{m_\circ c^2}{2\rho_\circ} [\gamma^\mu, \gamma^\nu] \psi\,
 \overline{\psi}J_{\mu\nu}\psi 
+
  i  \frac{ \ell e^2}{\hbar c}[\gamma^\mu, \gamma^\nu] \psi
  \left(
 - \frac{m_\circ c^2 }{e^2\rho_\circ}  e \overline{\psi}
\gamma_\mu  \psi
\right)
A_\nu
&=
0
\\
 g^{\mu\nu} e \overline{\psi}\gamma_\mu \psi
&=
\partial_\mu F^{\mu\nu}
.
\end{align}
\end{subequations}
A superconducting fluid (with magnetized quantum vortices) has the solution 
\begin{align}
\label{SM_superconducting_fluid_generalized_London_relation_ansatz}
   A_\mu 
   &= - \frac{m_\circ c^2 }{e^2\rho_\circ}  e \overline{\psi}
\gamma_\mu  \psi
,
\end{align}
so  (\ref{SM_Euler_Lagrange_equations_nonlinear_GFT_5}) in this case becomes
\begin{subequations}
\label{SM_Euler_Lagrange_equations_nonlinear_GFT_5}
\begin{align}
i\hbar c g^{\mu\nu}\gamma_\mu \left(\partial_\nu  - i \frac{eA_\nu}{\hbar c}\right) \psi
+
i \frac{m_\circ c^2}{2\rho_\circ} [\gamma^\mu, \gamma^\nu] \psi\,
 \overline{\psi}J_{\mu\nu}\psi 
+
  i  \frac{ \ell e^2}{\hbar c}[\gamma^\mu, \gamma^\nu] \psi
A_\mu
A_\nu
&=
0
\\
 g^{\mu\nu} e \overline{\psi}\gamma_\mu \psi
&=
\partial_\mu F^{\mu\nu}
.
\end{align}
\end{subequations}
   \end{widetext}
Since the commutator  $[\gamma^\mu, \gamma^\nu]$ is antisymmetric and $A_\mu A_\nu$ is symmetric in $\mu$ and $\nu$, so the last term in the first equation cancels to zero and the equations of motion become simpler in this case
\begin{subequations}
\label{SM_Euler_Lagrange_equations_nonlinear_GFT_6}
\begin{align}
i\hbar c g^{\mu\nu}\gamma_\mu \left(\partial_\nu  - i \frac{eA_\nu}{\hbar c}\right) \psi
+
i \frac{m_\circ c^2}{2\rho_\circ} [\gamma^\mu, \gamma^\nu] \psi\,
 \overline{\psi}J_{\mu\nu}\psi 
&=
0
\\
 g^{\mu\nu} e \overline{\psi}\gamma_\mu \psi
&=
\partial_\mu F^{\mu\nu}
.
\end{align}
\end{subequations}
The  generator of angular momentum in the spin-representation  of the Lorentz group is
\begin{align}
   S^{\mu\nu} &= 
   \frac{1}{4}[\gamma^\mu, \gamma^\nu] 
\end{align}
 where the spin generator satisfies the algebra 
$[S^{\mu \nu},S^{\rho \sigma}] =
 i\left(\eta^{\nu \rho}S^{\mu \sigma}-\eta^{\rho \mu}S^{\nu \sigma}-\eta^{\nu\sigma}S^{\mu\rho}+\eta^{\mu\sigma}S^{\nu\rho}\right)$. 
So the equations of motion may be written as
\begin{subequations}
\label{SM_Euler_Lagrange_equations_nonlinear_GFT_7}
\begin{align}
i\hbar c g^{\mu\nu}\gamma_\mu \left(\partial_\nu  - i \frac{eA_\nu}{\hbar c}\right) \psi
&+
i \frac{2m_\circ c^2}{\rho_\circ} S^{\mu\nu} \psi\,
 \overline{\psi}J_{\mu\nu}\psi 
=
0
\\
 g^{\mu\nu} e \overline{\psi}\gamma_\mu \psi
&=
\partial_\mu F^{\mu\nu}
.
\end{align}
\end{subequations}
These are the Euler-Lagrange equation presented in the {\it A curved-space gauge theory} section of the Letter.

\vspace{1em}

\subsubsection{Relativistic superconductivity}

   \begin{widetext}

Using (\ref{SM_torsional_space_metric_tensor_field_Smunu_form}), this may be separated as
\begin{subequations}
\label{SM_Euler_Lagrange_equations_nonlinear_GFT_8}
\begin{align}
i\hbar c \eta^{\mu\nu}\gamma_\mu \left(\partial_\nu  - i \frac{eA_\nu}{\hbar c}\right) \psi
+
i\hbar c  \frac{ \overline{\psi}S^{\mu\nu} \psi}{\overline{\psi} \psi}
\gamma_\mu \left(\partial_\nu  - i \frac{eA_\nu}{\hbar c}\right) \psi
+
i \frac{2m_\circ c^2}{\rho_\circ} S^{\mu\nu} \psi\,
 \overline{\psi}J_{\mu\nu}\psi 
&=
0
\\
 g^{\mu\nu} e \overline{\psi}\gamma_\mu \psi
&=
\partial_\mu F^{\mu\nu}
.
\end{align}
\end{subequations}
Using (\ref{SM_flux_expansion}), this becomes
\begin{subequations}
\label{SM_Euler_Lagrange_equations_nonlinear_GFT_9}
\begin{align}
i\hbar c \eta^{\mu\nu}\gamma_\mu \left(\partial_\nu  - i \frac{eA_\nu}{\hbar c}\right) \psi
+
i\hbar c  \frac{ \overline{\psi}S^{\mu\nu} \psi}{\overline{\psi} \psi}
\gamma_\mu \left(\partial_\nu  - i \frac{eA_\nu}{\hbar c}\right) \psi
+
\frac{\hbar c}{2\ell}S^{\mu\nu} \psi\,
\frac{ \overline{\psi}J_{\mu\nu}\psi }{\overline{\psi} \psi}
&=
0
\\
 g^{\mu\nu} e \overline{\psi}\gamma_\mu \psi
&=
\partial_\mu F^{\mu\nu}
.
\end{align}
\end{subequations}
Using (\ref{SM_position_representation_angular_momentum_operator}), this becomes
\begin{subequations}
\label{SM_Euler_Lagrange_equations_nonlinear_GFT_10}
\begin{align}
\label{SM_Euler_Lagrange_equations_nonlinear_GFT_10_a}
i\hbar c \eta^{\mu\nu}\gamma_\mu \left(\partial_\nu  - i \frac{eA_\nu}{\hbar c}\right) \psi
+
\frac{\hbar c }{2\ell} \frac{ \overline{\psi}S^{\mu\nu} \psi}{\overline{\psi} \psi}
J_{\mu\nu} \psi
- 
\frac{ \overline{\psi}S^{\mu\nu} \psi}{\overline{\psi} \psi}
\gamma_\mu eA_\nu \psi
+
\frac{\hbar c}{2\ell}S^{\mu\nu} \psi\,
\frac{ \overline{\psi}J_{\mu\nu}\psi }{\overline{\psi} \psi}
&=
0
\\
 g^{\mu\nu} e \overline{\psi}\gamma_\mu \psi
&=
\partial_\mu F^{\mu\nu}
.
\end{align}
\end{subequations}
   \end{widetext}

  Ansatz (\ref{SM_torsional_space_metric_tensor_field_Smunu_form}) equates the spin-1/2 fermion's intrinsic spin  to the torsion in the space the fermion occupies. If one equates the expectation value of the fermion's angular momentum (in units of  $mc\ell$) to the expected value of the intrinsic spin (in units of $\hbar$) due to torsion, then the spin-torsion ansatz (\ref{SM_torsional_space_metric_tensor_field_Smunu_form}) may be expressed in a corollary form as  an angular momentum quantization condition    %
\begin{align}
\label{SM_spin_torsional_ansatz}
\hbar\langle
S^{\mu\nu} 
\rangle
=
-
mc\ell
\langle J^{\mu\nu} \rangle
  ,
\end{align}
  where the expectation value of an operator $\hat O$ is defined as 
\begin{align}
\label{SM_expected_value_definition}
     \langle \hat O \rangle \equiv \frac{\overline{\psi}\hat O \psi}{\overline{\psi}\psi}
  . 
\end{align}
The terms with a contraction of $S^{\mu\nu}$ with $J_{\mu\nu}$ on the lefthand side of (\ref{SM_Euler_Lagrange_equations_nonlinear_GFT_10_a}) 
may be reduced.  They are related to the product of expectation values
\begin{align}
      \frac{\hbar c}{\ell} 
      \langle
S^{\mu\nu}
\rangle
  \langle
  J_{\mu\nu}
  \rangle
\stackrel{(\ref{SM_spin_torsional_ansatz})}{=}
 -   mc^2
      ,
\end{align}
using the normalization condition  $\langle J^{\mu\nu} \rangle\langle J_{\mu\nu} \rangle =1$. This may be written in either of two  ways
\begin{subequations}
\begin{align}
      \frac{\hbar c}{2\ell} 
 \overline{\psi}S^{\mu\nu} \psi
  \langle
  J_{\mu\nu}
  \rangle
\stackrel{(\ref{SM_expected_value_definition})}{=}
 -   \frac{mc^2}{2}
  \overline{\psi}\psi
\end{align}
or
\begin{align}
      \frac{\hbar c}{2\ell} 
  \langle
  S_{\mu\nu}
  \rangle
 \overline{\psi}J_{\mu\nu} \psi
\stackrel{(\ref{SM_expected_value_definition})}{=}
 -   \frac{mc^2}{2}
  \overline{\psi}\psi
  ,
\end{align}
\end{subequations}
from which follows the operator-valued eigenequations
\begin{subequations}
\begin{align}
      \frac{\hbar c}{2\ell} 
 S^{\mu\nu} \psi
  \langle
  J_{\mu\nu}
  \rangle
=
 -   \frac{mc^2}{2}
\psi
\end{align}
or
\begin{align}
      \frac{\hbar c}{2\ell} 
  \langle
  S_{\mu\nu}
  \rangle
J_{\mu\nu} \psi
=
 -   \frac{mc^2}{2}
\psi
.
\end{align}
\end{subequations}
This implies that (\ref{SM_Euler_Lagrange_equations_nonlinear_GFT_10}) reduces to
\begin{subequations}
\label{SM_Euler_Lagrange_equations_nonlinear_GFT_11}
\begin{align}
\label{SM_Euler_Lagrange_equations_nonlinear_GFT_11_a}
i\hbar c \eta^{\mu\nu}\gamma_\mu \left(\partial_\nu  - i \frac{eA_\nu}{\hbar c}\right) \psi
-
 mc^2
\psi
&- 
\frac{ \overline{\psi}S^{\mu\nu} \psi}{\overline{\psi} \psi}
\gamma_\mu eA_\nu \psi
=
0
\\
 g^{\mu\nu} e \overline{\psi}\gamma_\mu \psi
&=
\partial_\mu F^{\mu\nu}
.
\end{align}
\end{subequations}
The last term on the lefthand side of (\ref{SM_Euler_Lagrange_equations_nonlinear_GFT_11_a}) is  related to
\begin{subequations}
\begin{align}
   \langle S^{\mu\nu}\rangle
   \overline{\psi}
\gamma_\mu eA_\nu \psi
&\stackrel{(\ref{SM_superconducting_fluid_generalized_London_relation_ansatz})}{=}
 - \frac{e^2\rho_\circ}{m_\circ c^2 }  \langle S^{\mu\nu}\rangle
A_\mu A_\nu 
\\
&\stackrel{(\ref{SM_flux_expansion})}{=}
\frac{4e^2}{\hbar c}
\langle S^{\mu\nu}\rangle
A_\mu A_\nu    \overline{\psi}\psi
,
\end{align}
\end{subequations}
from which follows from the eigenequation
\begin{align}
   \langle S^{\mu\nu}\rangle
\gamma_\mu eA_\nu \psi
=
\frac{4e^2}{\hbar c}
\langle S^{\mu\nu}\rangle
A_\mu A_\nu   
 \psi
 .
 \end{align}
Since the spin generator $S^{\mu\nu}$ is antisymmetric and $A_\mu A_\nu$ is symmetric in $\mu$ and $\nu$, the righthand side of this equation must vanish
\begin{align}
\label{SM_S_munu_gamma_mu_A_mu_psi_equals_zero}
   \langle S^{\mu\nu}\rangle
\gamma_\mu eA_\nu \psi
=
0 .
 \end{align}
Finally, inserting (\ref{SM_S_munu_gamma_mu_A_mu_psi_equals_zero}) into (\ref{SM_Euler_Lagrange_equations_nonlinear_GFT_11}) gives the equations of motion in  a much simpler form
\begin{subequations}
\label{SM_Euler_Lagrange_equations_nonlinear_GFT_12}
\begin{align}
i\hbar c \gamma^\nu \left(\partial_\nu  - i \frac{eA_\nu}{\hbar c}\right) \psi
-
 mc^2
\psi
&=
0
\\
 g^{\mu\nu} e \overline{\psi}\gamma_\mu \psi
&=
\partial_\mu F^{\mu\nu}
\\
\label{SM_Yepez_Dirac_Maxwell_London_equations_of_motion_c_lower_bound}
F^{\mu\nu} &=\partial^\mu A^\nu - \partial^\nu A^\mu
,
\end{align}
\end{subequations}
where the definition of the field strength tensor  (\ref{SM_Yepez_Dirac_Maxwell_London_equations_of_motion_c_lower_bound}) is added as a component equation. 
These are the equations of motion for a superconducting Fermi fluid that are presented in the {\it Relativistic superconductivity} section of the Letter.

\subsection{Outgoing 4-potential}

The outgoing fermion field $\psi'$ to first order is
\begin{equation}
\label{SM_psi_A_forward_reaction_expansion}
\psi'(x)
=
\psi(x) 
-
i\ell  \gamma_\mu \frac{e A^\mu(x)}{\hbar c} \psi(x)
+
\cdots.
\end{equation}
Then, upon making use of the adjoint gamma matrices $\gamma^{\mu\dagger}$ and the anticommutation relation $\{\gamma^{\mu\dagger}, \gamma^{\nu\dagger}\}=2\eta^{\mu\nu}$,  an expansion for the outgoing probability current density $J'^\nu(x)=\overline{\psi'}(x)  \gamma^\nu \psi'(x)$ is
   \begin{widetext}
\begin{subequations}
\label{SM_Outgoing_current_density_expansion}
\begin{eqnarray}
J'^\nu(x)
&=&
\psi'^\dagger(x)\gamma_0  \gamma^\nu \psi'(x)
\\
&\stackrel{(\ref{SM_psi_A_forward_reaction_expansion})}{=}&
\left(
\psi^\dagger(x) 
+
i\ell  \frac{e A^\mu(x)}{\hbar c} \psi^\dagger(x) \gamma_\mu^\dagger 
+
\cdots
\right)
\gamma_0  \gamma^\nu
\left(
\psi(x) 
-
i\ell  \gamma_\kappa \frac{e A^\kappa(x)}{\hbar c} \psi(x)
+
\cdots
\right)
\\
&=&
\psi^\dagger(x)\gamma_0  \gamma^\nu \psi(x)
+
i\ell  \frac{e A^\mu(x)}{\hbar c} \psi^\dagger(x)
 \gamma_\mu^\dagger 
\gamma_0  \gamma^\nu
\psi(x) 
-
i\ell 
  \frac{e A^\kappa(x)}{\hbar c} 
\psi^\dagger(x) 
\gamma_0  \gamma^\nu
 \gamma_\kappa
  \psi(x)
+
\cdots
\\
&=&
J^\nu(x)
+
\left.
i\ell 
  \frac{e A^\mu(x)}{\hbar c} 
\psi^\dagger(x) 
\middle(
 \gamma_\mu^\dagger
\gamma_0  
\gamma^\nu
-
\gamma_0  
\gamma^\nu
 \gamma_\mu
\right)
  \psi(x)
+
\cdots
\\
&=&
J^\nu(x)
+
\left.
i\ell 
  \frac{e A_\mu(x)}{\hbar c} 
\psi^\dagger(x) 
\middle(
2\eta^{0\mu}
\gamma^\nu
-
\gamma_0 
 \gamma^{\mu\dagger}
 \gamma^\nu
-
\gamma_0  
\gamma^\nu
 \gamma^\mu 
\right)
  \psi(x)
+
\cdots
\\
\label{SM_Outgoing_current_density_expansion_f}
&=&
J^\nu(x)
+
\left.
i\ell 
  \frac{e A_\mu(x)}{\hbar c} 
\psi^\dagger(x) 
\middle(
2\eta^{0\mu}
\gamma^\nu
-
\gamma_0
\left(
 \gamma^{\mu\dagger} 
 \gamma^\nu
+
  \gamma^\mu \gamma^\nu
  -
  [ \gamma^{\mu},  \gamma^\nu]
 \right)
 \right)
  \psi(x)
+
\cdots
\\
\label{SM_Outgoing_current_density_expansion_g}
&=&
J^\nu(x)
+
\left.
i\ell 
  \frac{e A_\mu(x)}{\hbar c} 
\psi^\dagger(x) 
\middle(
\left(
2\eta^{0\mu}
-
 \gamma_0
 \gamma^{\mu\dagger} 
 -
  \gamma_0
  \gamma^\mu 
\right)
\gamma^\nu
-
\gamma_0
  [ \gamma^{\mu},  \gamma^\nu]
  \right)
  \psi(x)
+
\cdots
\\
\label{SM_Outgoing_current_density_expansion_h}
&=&
J^\nu(x)
-
i\ell 
  \frac{e A_\mu(x)}{\hbar c} 
\psi^\dagger(x) \gamma_0
  [ \gamma^{\mu},  \gamma^\nu]
  \psi(x)
+
\cdots.
\end{eqnarray}
\end{subequations}
Finally, with the  identity $A^\nu(x) = - e \lambda_\text{\tiny L}^2 J^\nu(x) = -\frac{m_\circ c^2}{e\rho_\circ}J^\nu(x)$, an analytical expansion of the back reaction is obtained
\begin{subequations}
\label{SM_psi_A_backward_reaction_expansion}
\begin{eqnarray}
\label{SM_psi_A_backward_reaction_expansion_a}
A'^\nu(x)
&=&
A^\nu(x)
+
i 
  \frac{  m_\circ c^2\ell }{\hbar c \rho_\circ} 
{\psi}^\dagger(x)
 \gamma_0
[ \gamma^\mu ,
 \gamma^\nu
 ]
  \psi(x)
A_\mu(x)
+
\cdots
\end{eqnarray}
\end{subequations}
   \end{widetext}
This is the back-reaction equation expansion presented in the  {\it Forward and back reactions} section of the Letter.  This derivation originally appeared in Ref.~\cite{yepez_arXiv1609.02225v2_quant_ph}. 
Here the outgoing  local value of the Maxwell field $A'^\nu(x)$ is determined by the local values of the incoming Dirac field $\psi^\dagger(x)$, the outgoing Dirac field $\psi(x)$, and the incoming value of the Maxwell field $A^\nu(x)$.

\subsection{Maxwell equations in spinor form}

Let us first rewrite the Maxwell equations
\begin{subequations}
\label{SM_Maxwell_equations}
\begin{align}
\label{SM_Maxwell_equations_a}
\eta^{\mu\nu} e \overline{\psi}\gamma_\mu \psi
&=
\partial_\mu F^{\mu\nu}
\\
\label{SM_Maxwell_equations_b}
F^{\mu\nu} &=\partial^\mu A^\nu - \partial^\nu A^\mu
\end{align}
\end{subequations}
 in spinor form. With real-valued  a charged 4-current source field $e J^\nu = e(\rho, \bm{J})=e \overline{\psi}\gamma^\nu \psi$,  number density $\rho$ and 3-current $\bm{J}$, (\ref{SM_Maxwell_equations})  may be written in term of  a complex 3-vector field $\bm{F}$
\begin{subequations}
\label{SM_Maxwell_equations_complex_3_vector_form}
\begin{align}
\label{SM_Maxwell_equations_complex_3_vector_form_a}
   \nabla\cdot \bm{F}  &=  e \rho,
   \qquad
   i \partial_t \bm{F}  =  
\nabla\times\bm{F} - i e \bm{J} 
\\
\label{SM_Maxwell_equations_complex_3_vector_form_b}
\bm{F} &= - \partial_0 \bm{A} -\nabla A_0 + i \nabla\times \bm{A}
.
\end{align}
\end{subequations}
%
 %
To combine  
(\ref{SM_Maxwell_equations_complex_3_vector_form_a}) into a single equation, one may use a novel complex 4-vector field $F^\mu = (F_0, \bm{F})$, so (\ref{SM_Maxwell_equations_complex_3_vector_form}) becomes
\begin{subequations}
\label{SM_Maxwell_equations_complex_valued_4_vector_form}
\begin{align}
\label{SM_Maxwell_equations_2}
e J^\nu&=( \nabla \cdot \bm{F}, -\partial_0 \bm{F} - i \nabla\times \bm{F})
\\
\label{SM_Maxwell_equations_1}
F^\mu&=(-\partial\cdot A, -\partial_0\bm{A}-\partial_0 A_0+ i \nabla\times \bm{A})
,
\end{align}
\end{subequations}
which can be written elegantly in spinor variables.

Let us start by converting $A^\mu=(A_0, A_x, A_y, A_z)^\text{T}$
  into the 4-spinor field, say ${\cal A}$, by using a unitary matrix transformation, say $T$.  The unitary transformation is ${\cal A}_a =  T_{a\mu} A^\mu$, which component form is
 \begin{align}
\label{SM_4_spinor_A}
 {\scriptsize
\begin{pmatrix}
      {\cal A}_{\text{\tiny L}\uparrow}    \\ 
      {\cal A}_{\text{\tiny L}\downarrow}    \\ 
      {\cal A}_{\text{\tiny R}\uparrow}    \\ 
      {\cal A}_{\text{\tiny R}\downarrow}    \\ 
\end{pmatrix}
}
=
  {\scriptsize
 \underbrace{
\frac{1}{\sqrt{2}}
  \begin{pmatrix}
  0    &   -1 & i & 0 \\
 1     &  0 & 0 & 1 \\
 -1 & 0 & 0 & 1\\
 0 & 1 & i & 0 \\ 
\end{pmatrix}
}_{T}
\begin{pmatrix}
      A_0    \\
      A_x    \\
      A_y    \\
      A_z    \\
\end{pmatrix}
}
=
 \frac{1}{\sqrt{2}}
 {\scriptsize
\begin{pmatrix}
      -A_x + i A_y    \\
  A_0+   A_z \\
   -A_0+   A_z\\
     A_x + i A_y  \\
\end{pmatrix}
}
.
\end{align}
Similarly, along with the 4-spinor potential  field ${\cal A}$, denote 
the  4-spinor electromagnetic  field $\tilde {\cal F}$,   current density  field ${\cal J}$,  and  dual 4-potential field $\tilde {\cal A}$  respectively as
\begin{equation}
\label{SM_4_spinor_field_definititions}
\begin{split}
{\cal A} 
& =  
 \frac{1}{\sqrt{2}}
 {\scriptsize
\begin{pmatrix}
      -A_x + i A_y    \\
  A_0+   A_z \\
   -A_0+   A_z\\
     A_x + i A_y  \\
\end{pmatrix}
},
\quad
\tilde {\cal F }
 =  
 \frac{1}{\sqrt{2}}
 {\scriptsize
\begin{pmatrix}
     - F_x + i F_y    \\
       - \partial\cdot A  + F_z \\
        \partial\cdot A + F_z \\
      F_x + i F_y  
\end{pmatrix}
}
\\
{\cal J }
 &= 
 \frac{1}{\sqrt{2}}
 {\scriptsize
\begin{pmatrix}
      -J_x + i J_y    \\
   \rho  +J_z\\
    -\rho +J_z\\
     J_x + i J_y  \\
\end{pmatrix}
}
,
\quad
\tilde {\cal A}
=
-\frac{i \lambda_\text{\tiny L}}{ \sqrt{2}}
 {\scriptsize
\begin{pmatrix}
     - F_x + i F_y    \\
       - \partial\cdot A  + F_z \\
        \partial\cdot A + F_z \\
      F_x + i F_y  
\end{pmatrix}
}
.
\end{split}
\end{equation}

The Maxwell equations (\ref{SM_Maxwell_equations_complex_valued_4_vector_form})  expressed in terms of 4-spinor fields (\ref{SM_4_spinor_field_definititions}) and using tensor-product notation  are \cite{yepez_arXiv1609.02225v2_quant_ph}
\begin{align}
\label{SM_Maxwell_equation_4spinor_rep_covariant_form}
e {\cal J} +  \bm{1}\otimes \sigma\cdot \partial \tilde {\cal F} &=
0,
\qquad
\tilde {\cal F} +  \bm{1}\otimes   \bar\sigma\cdot \partial {\cal A}=
0,
\end{align}
where
$\sigma^\mu = (1,\bm{\sigma})$, $\bar\sigma^\mu=(1,-\bm{\sigma})$.
These are the spinor form of the Maxwell equations presented in the  {\it Quantum computational spinor form} section of the Letter. 

\subsection{Maxwell-London equations}

 The relativistic London relation
  $e J_\mu(x)=-A_\mu(x)/
\lambda_\text{\tiny $L$}^2$ 
  in the spinor variables reduces to $e {\cal J}=-{\cal A}/ \lambda_\text{\tiny L}^2$,  and since $\tilde {\cal F} \equiv i  \tilde {\cal A}/\lambda_\text{\tiny L}$ in (\ref{SM_4_spinor_field_definititions}),
  (\ref{SM_Maxwell_equation_4spinor_rep_covariant_form}) reduce to   
%
 a pair of spinor equations coupling ${\cal A}$ to $\tilde {\cal A}$
\begin{align}
\label{SM_Maxwell_equation_4spinor_rep_covariant_form_A_tildeA}
-{\cal A}/ \lambda_\text{\tiny L}+ i \bm{1}\otimes \sigma\cdot \partial  \tilde {\cal A} &=
0,
\quad
 -  \tilde {\cal A}/\lambda_\text{\tiny L}+ i \bm{1}\otimes   \bar\sigma\cdot \partial {\cal A}=
0.
\end{align}
  Therefore, 
  the Dirac equation $i \hbar c  \gamma^\nu
\Big(
\partial_{\nu}+\frac{ieA_\nu}{\hbar c}
\Big)
\psi
-
 m c^2 \psi
=0$
  and (\ref{SM_Maxwell_equation_4spinor_rep_covariant_form_A_tildeA}) are the equation of motion for a superconducting Fermi condensate, which in component form  using single  and doublet spinors are 
\begin{subequations}
\label{SM_superconductivity_equation_spinor_form}
\begin{align}
\begin{pmatrix}
-\frac{mc}{\hbar}
    &  i \sigma \cdot \left(\partial +i \frac{e A}{\hbar c}\right) \\
  i \bar{\sigma} \cdot \left(\partial+i\frac{eA}{\hbar c}\right)    &
-\frac{mc}{\hbar}
\end{pmatrix}
\begin{pmatrix}
      \psi_\text{\tiny L}    \\
       \psi_\text{\tiny R}  
\end{pmatrix}
&=
 0
\\
\label{SM_superconductivity_equations_doublet_form}
\begin{pmatrix}
  - \frac{1}{\lambda_\text{\tiny $L$}}   &  i \bm{1}\otimes\sigma \cdot \partial \\
  i  \bm{1}\otimes\bar{\sigma} \cdot \partial    &  - \frac{1}{\lambda_\text{\tiny $L$}}
\end{pmatrix}
\begin{pmatrix}
      {\cal A}    \\
     \tilde {\cal A}
\end{pmatrix}
&=
 0.
\end{align}
\end{subequations}
Equations (\ref{SM_superconductivity_equation_spinor_form}) are a  spinor representation of the relativistic Dirac-Maxwell-London equations \cite{yepez_arXiv1609.02225v2_quant_ph}---(\ref{SM_superconductivity_equations_doublet_form}) is a manifestly covariant and symmetrical Dirac equation  for the doublet  field $({\cal A}, \tilde{\cal A})^\text{T}$. 
Symmetry hidden in superconducting electrodynamics is revealed when a mass parameter $\lambda_\text{\tiny L}^{-1}=m_\text{\tiny L}c/\hbar$ is used in the gauge field dynamics, a feature of a relativistic superconducting fluid.  
A generalization of these are the equations of motion for a relativistic superconducting Fermi fluid are presented in the  {\it Quantum computational spinor form} section of the Letter.

\subsection{Quantum algorithm}

\subsubsection{Path integration}
The block diagonal mass matrix is
\begin{subequations}
\begin{align}
\label{SM_M_L_formula}
  M_\text{\tiny $L$}(x) 
    &\equiv
        \frac{\hbar}{\lambda_\text{\tiny $L$} c} 
    \begin{pmatrix}
1
-
 \frac{T M(x) T^\dagger c \ell}{\hbar}
 &
 0
      \\
0&    1 
\end{pmatrix}
\\
&=
        \frac{\hbar}{\lambda_\text{\tiny $L$} c} 
-
\frac{\ell}{\lambda_\text{\tiny $L$}}
    \begin{pmatrix}
T M(x) T^\dagger
 &
 0
      \\
0&    0 
\end{pmatrix}
,
\end{align}
\end{subequations}
where the unitary transformation  $T_{a\mu}$  is given in (\ref{SM_4_spinor_A})
\begin{align}
T_{a\mu}
=
{\scriptsize
\frac{1}{\sqrt{2}}
  \begin{pmatrix}
  0    &   -1 & i & 0 \\
 1     &  0 & 0 & 1 \\
 -1 & 0 & 0 & 1\\
 0 & 1 & i & 0 \\ 
\end{pmatrix}
}_{a\mu}
\end{align}
and
\begin{align}
 {M^\nu}_\mu(x) = i m_\circ 
   \frac{\overline{\psi}(x)
[\gamma^\nu,\gamma_\mu]
  \psi(x)}{\rho_\circ}
  .
  \end{align}
With the Dirac matrices $\bm{\alpha} = \sigma_x \bm{\sigma}$ and $\beta = \sigma_z \bm{1}$,  the contravariant 4-vector of gamma matrices are
\begin{equation}
\gamma^\mu = (\gamma^0, \bm{\gamma}) = (\sigma_z\bm{1},  i \sigma_y \bm{\sigma}).
\end{equation}
The local equilibrium conditions  (that hold at all points of the spacetime lattice) 
\begin{subequations}
\label{SM_QID_equations_symmetrical_unitary_form}
\begin{align}
\psi(x)
 &=
e^{- \ell  \gamma \cdot  
\left(\partial
 + i \frac{e  A(x) }{\hbar c}
 \right) 
- i  \frac{  m c^2 \tau}{\hbar}  
} 
\psi(x)
\label{SM_QID_equations_symmetrical_unitary_form_psi_evolution}
    \\
\Phi(x)
 &=
e^{-\ell  {\cal G}
 \cdot 
 \partial
- i  \frac{  M_\text{\tiny $L$}(x) c^2 \tau}{\hbar}
}\Phi(x) 
,
\label{SM_QID_equations_symmetrical_unitary_form_Phi_evolution}
\end{align}
\end{subequations}
  can be recast as unitary evolution of  a composite fermionic  field $\Psi\equiv (\psi, \Phi)$ and emulated with a quantum lattice gas algorithm for relativistic quantum mechanics \cite{yepez_arXiv1512.02550_quant_ph}, which is based on a  path summation rule on a spacetime lattice.
To implement 
theory (\ref{SM_nonlinear_GFT_lagrangian_density}) in path integral form
\begin{align}
\label{SM_standard_path_integral}
K_{ab} 
\equiv
\langle
 \hat{K}_{ab}
\rangle
= 
\int_a^b{\cal D}\{ x\} \,
\int\frac{dp^4}{(2\pi\hbar)^4}
e^{\frac{i}{\hbar} \int d^4x \,{\cal L}  }
\end{align}
on a spacetime lattice of size $T L^3$,  a set of spin chains $\{{s}^\mu_0,\dots,{s}^\mu_{N-1}\}$   enumerate the paths of the fermions, where the constant  magnetization $\ell\sum_{w=0}^{N-1} {s}^\mu_w
= {x}^\mu_b-{x}^\mu_a$ in spin space 
correspondes to fixed endpoints $x^\mu_a$ and ${x}^\mu_b$ in spacetime.  The fermion's 4-momentum ${p}^\mu_{n} \equiv
 \frac{2\pi \hbar}{\ell}\left( \frac{n_t}{T}, \frac{n_x}{L}, \frac{n_y}{L}, \frac{n_z}{L}\right)$ is represented in reciprocal space by integers $n=(n_t, n_x, n_y, n_z)$, and the 4-momentum integral is represented by a summation in  reciprocal space 
 $
\sum_{{n}}\equiv \sum_{n_{t} = -T/2}^{(T/2)-1}
\sum_{n_{x} = -L/2}^{(L/2)-1} \sum_{n_{y} = -L/2}^{(L/2)-1} \sum_{n_{z} = -L/2}^{(L/2)-1}$. 
An extra 4-spinor field  $\tilde \psi$ is added for the purpose of introducing a  16-component (4-qubit \cite{yepez_arXiv1609.02225v2_quant_ph}) multiplet field $\Psi = ({\cal A}, \tilde {\cal A}, \psi, \tilde \psi)$.  
 To this end, it is convenient to use  $\Delta^\mu = (\Delta_0, \bm{\Delta})$, with Dirac matrices 
$
\Delta^0
\equiv 
n\otimes\bm{1}\otimes\gamma^0
+
h\otimes{\cal G}^0$ and $\bm{\Delta}
\equiv
 n\otimes\bm{1}\otimes \bm{\gamma}
+
 h\otimes \bm{{\cal G}}$,
and with a 4-potential $G^\mu=(G_0, \bm{G})
 =  
n\otimes h \otimes \bm{1}_4 \, e A^\mu$, 
where $n=(1-\sigma_z)/2$ is the singleton qubit number  operator  and $h=1-n$ is the hole operator.  The  Dirac Hamiltonian 
${\hat{h}_\text{\tiny D}}
 =
  \Delta^0
\bm{\Delta} \cdot
\left(
 {\bm{\hat{p}}}_{\bm{n}} c 
 -  \bm{G}
 \right)
  +   \Delta^0 {m}_\text{\tiny HE}c^2$  generates the particle and field dynamics, 
where   the block diagonal mass matrix is ${m}_\text{\tiny HE}\equiv n\otimes \bm{1}_8\,  {m} + h\otimes \bm{1} \otimes  {M_\text{\tiny $L$}} $.  So to model (\ref{SM_nonlinear_GFT_lagrangian_density})
 on a qubit array, one can use a highest-energy (HE) kernel operator $\hat{K}^\text{\tiny HE}_{ab}$, where  $\hat{K}^\text{\tiny HE}_{ab}\rightarrow \hat{K}_{ab}$ in the small $\ell$ limit.  The highest-energy kernel operator  is

\begin{equation}
\label{SM_yepez_feynman_path_summation_spin_form4}
\hat{K}^\text{\tiny HE}_{ab}
\!
=
\!\!
\sum_{{n}}
\frac{1}{(\ell T)}\frac{1}{(\ell L)^3}
e^{-i{x_\mu p^\mu_{n}}/{\hbar}}
\!
\sum_{\text{paths}}
\!\!
\ell^3 
e^{-\frac{i}{\hbar}  \sum_{w=0}^{N-1} \delta t \,{(E'-\hat{h}_\text{\tiny D}}) },
\end{equation}
where $E'= E-G_0$  and $\delta t= \zeta \tau$ is the time differential for scalar time-scale factor $\zeta$ \cite{yepez_arXiv1512.02550_quant_ph}.   At the highest-energy scale, (\ref{SM_yepez_feynman_path_summation_spin_form4}) leads to  an equation of motion in  unitary  form
\begin{equation}
\label{SM_HE_unitary_EoM_for_Psi}
\Psi(x)
 =
e^{-\ell  \Delta
 \cdot 
\left(
 \partial
 + i \frac{G(x)}{\hbar c}
 \right)
- i  {{m}_\text{\tiny HE} c\ell}/{\hbar}
}\Psi(x) ,
\end{equation}
equivalent to  (\ref{SM_QID_equations_symmetrical_unitary_form}) 
and the forward and back  reactions are  expressed as a single  reaction
$\Psi'(x)
 = 
e^{-i\ell  \Delta_\mu \frac{G^\mu(x)}{\hbar c}}\Psi(x)$. 

 With  stream operator ${\hat{\cal S}}(\bm{n}) 
=e^{ i\ell \sigma_z
 \bm{\sigma}\cdot \hat{\bm{p}}_{\bm{n}} /\hbar
}$
expressed with the momentum operator $\hat{\bm{p}}_{\bm{n}}  = {\hbar \bm{n}\cdot\bm{\sigma}}/{(L\ell)}$,  one converts to the Bloch-Wannier picture by replacing the  spacetime operators $\hat{\bm{p}}_{\bm{n}}$ and $\hat{E}_{\bm{n}}$  on the qubit array with derivative operators acting on a  Dirac field $\Psi(x)$ in continuous spacetime \cite{yepez_arXiv1512.02550_quant_ph}, i.e.
$\bm{\hat{p}}_{\bm{n}} \sim -i\hbar\nabla$ 
 and 
$\hat{E}_{\bm{n}} \sim i\hbar\partial_t$. 
In this way,  (\ref{SM_yepez_feynman_path_summation_spin_form4})     is equivalent to a path integral for an action operator 
\begin{subequations}
\label{SM_feynman_path_integral_in_p_space}
\begin{align}
\hat{K}^\text{\tiny HE}_{ab}
&\cong
\int_a^b {\cal D}\{ x\} 
\int\frac{dp^4}{(2\pi\hbar)^4}
\,
e^{\frac{i}{\hbar} \int dt\, {\hat{L}}},
\end{align}
\end{subequations}
where  the Lagrangian operator is ${\hat{L}  } =\hat{E}_{\bm{n}}- ( {\hat{h}_\text{\tiny D}}+ G_0)$ \cite{yepez_arXiv1512.02550_quant_ph}. %
The Lagrangian density 
(\ref{SM_nonlinear_GFT_lagrangian_density}) 
derives from the Lagrangian operator  ${\cal L}  
\equiv
\Psi^\dagger {\hat{L}  } 
\Psi$, which  is  concisely written as
${\cal L}  
=
i \hbar c \overline{\Psi}\, \Delta_\mu\left(
 \partial^{\mu}
 +\frac{i G^\mu}{\hbar c}
 \right) \Psi   -\overline{\Psi}  {m}_\text{\tiny HE}c^2  \Psi$,  
where the matrix element is congruent to the path integral in (\ref{SM_standard_path_integral}),  i.e. 
$\langle \hat{K}^\text{\tiny HE}_{ab}
\rangle
\cong 
\langle \hat{K}_{ab}
\rangle
$. 
The small $\ell$ expansion of (\ref{SM_HE_unitary_EoM_for_Psi}) are the Euler-Lagrange equation obtained
 by minimizing the action upon variation of $\Psi$.

\subsubsection{Stream-collide algorithm}

In the quantum lattice gas algorithm, chiral particle motion in the background 4-potential $G^\mu$ field is implemented by the stream operator
\begin{subequations}
\label{SM_QID_quantum_algorithm}
\begin{equation}
\label{SM_grand_stream_operator}
 {\cal S}(\bm{G}) 
 \approx  
 {\cal S}_ x(\bm{G}) {\cal S}_y(\bm{G}) {\cal S}_z(\bm{G}),
\end{equation}
where the stream operators along the $i$th direction are
${\cal S}_i(\bm{G})
=
e^{\ell  \Delta_0 \Delta_i 
 \partial_i}
e^{ i \ell \Delta_0 \Delta_i {G_i}/{(\hbar c)}
}$
 for $i=x,y,z$. The operator splitting (\ref{SM_grand_stream_operator}) can be implemented with error terms at fourth order, leading to a quantum algorithm that is numerically convergent to any arbitrary level of precision \cite{ yepez-qip-05}. Chiral symmetry is broken  by the collide operator
\begin{equation}
\label{SM_grand_collision_operator_in_the_QFT_limit}
 {\cal C} ( {m}_\text{\tiny HE})
 =
  \sqrt{1-\frac{ {m}_\text{\tiny HE}^2 c^4\tau^2}{\hbar^2}}
+
i
 \frac{{m}_\text{\tiny HE}c^2\tau}{\hbar}
\Delta_0
.
\end{equation}
The unitary evolution  (spacetime transfer) operator is
\begin{equation}
\label{SM_general_transfer_operator_highest_energy_quantum_algorithm_form}
{\hat{\cal U}}({n})
\!
= 
\!
e^{-i\frac{(E-G_0) \tau}{\hbar}} 
\!\!
\left[
\!
\sqrt{1-\frac{{m}_\text{\tiny HE}^2c^4\tau^2}{\hbar^2}}
S(\bm{G})
-
i
 \frac{{m}_\text{\tiny HE}c^2\tau}{\hbar}
\Delta_0
\right]
\!
,
\end{equation}
\end{subequations}
where $G_0(x)$ causes an overall phase rotation and 
 $\bm{G}(x)$  causes a direction-dependent phase rotation in the stream operator ${\cal S}(\bm{G})$ \cite{yepez_arXiv1512.02550_quant_ph}.\footnote{It is also possible to implement ${\cal S}(\bm{G})$ directly without operator splitting by performing a quantum Fourier transformation prior to streaming and performing an inverse quantum Fourier transformation after streaming.  These Fourier transformation are prescribed because the collide operator (\ref{SM_grand_collision_operator_in_the_QFT_limit}) is readily implemented in position space while the stream operator ${\cal S}(\bm{G})$  is readily implemented in momentum space.}   
   The update equation 
\begin{align}
\label{SM_nonlinear_gauge_field_theory_quantum_algorithm}
   \Psi(\bm{x}, t+ \tau)
=
{\hat{\cal U}}({n})
  \Psi(\bm{x},t)
\end{align}
is a  quantum algorithm for the many-body fermion system with particle-gauge field interactions; 
the evolution equation (\ref{SM_nonlinear_gauge_field_theory_quantum_algorithm})  is the quantum algorithm based on (\ref{SM_QID_equations_symmetrical_unitary_form}) as mentioned in the  {\it Quantum computational spinor form} section of the Letter. 
 This completes the quantum information dynamics representation of  many fermions interacting via a gauge field.

\end{document}